\documentclass[cameraready]{Interspeech}
\usepackage{xcolor}

\title{Towards Prosodically Informed Mizo TTS without Explicit Tone Markings}

\author[affiliation={1}, orcid=0000-0002-6368-0328] {Abhijit}{Mohanta}

\author[affiliation={2}]{Remruatpuii}{}

\author[affiliation={1}, orcid=0000-0002-9051-1255]{Priyankoo}{Sarmah}

\author[affiliation={1},orcid=0000-0002-0419-6501]{Rohit}{Sinha}

\author[affiliation={3},orcid=0000-0002-5419-5371]{Wendy}{Lalhminghlui}

\address{
    $^1$ Indian Institute of Technology Guwahati, India\\
    $^2$ Higher and Technical Institute Mizoram, India\\
    $^3$ University of Bern, Switzerland
}

\email{a.mohanta@iitg.ac.in, remruatpuii@hatim.ac.in, \{priyankoo, rsinha\}@iitg.ac.in, wendy.lalhminghlui@unibe.ch}


\usepackage{comment}

\begin{document}

\maketitle

\begin{abstract}
This paper reports on the development of a text-to-speech (TTS) system for Mizo, a low-resource tonal language of the Tibeto-Burman family, spoken primarily in the Indian state of Mizoram. The TTS was built using only 5.18 hours of data; however, based on subjective and objective evaluations, the outputs were considered perceptually acceptable and intelligible. A baseline Tacotron2 model was built, and then, using the same data, another TTS model was built with VITS. VITS achieved significantly lower Mel-Cepstral Distortion (MCD) and significantly higher subjective ratings than Tacotron2, although the differences in DNSMOS, F0 RMSE, and F0 correlation were not statistically significant.  In terms of tone synthesis, the VITS model showed numerically lower tone errors than the Tacotron2 model. The paper demonstrates that a non-autoregressive, end-to-end framework can achieve synthesis of acceptable perceptual quality and intelligibility.

\end{abstract}

\section{Introduction}

This paper describes an Mizo text-to-speech (TTS) synthesis system built using the Variational Inference with Adversarial Learning for Text-to-Speech (VITS) framework. VITS is an end-to-end (E2E) speech synthesis framework that is effective in low-resource settings. The paper describes the process of building the TTS system for Mizo, beginning with database collection and concluding with qualitative and quantitative evaluation. While TTS for high-resource languages has begun to achieve near-human levels of naturalness and intelligibility, it remains in its nascent stage for low-resource languages.

With the advent of deep neural network (DNN)-based TTS systems, the quality of the synthesized speech sounds has improved immensely \cite{wavnet}. These systems, such as WaveNet, consistently outperformed the unit selection-based concatenative and statistical parametric speech synthesis in terms of accuracy and naturalness. However, these initial systems required large amounts of audio data, and even synthesis demanded both time and greater computational power due to their autoregressive framework. The subsequent version, WaveNet2, could result in faster inference. Both versions of WaveNet were completely end-to-end (E2E), generating waveforms directly without an explicit vocoder. Following WaveNet, the Tacotron and Tacotron2 frameworks were proposed; they are also E2E frameworks but produce mel-spectrograms as outputs, requiring the explicit use of a vocoder \cite{tacotron1}. 

The non-autoregressive frameworks gained popularity in the following years due to their low inference time and ability to generate high-quality outputs with less data. The FastSpeech and FastSpeech2 frameworks produce mel-spectrograms as outputs; however, the latter is considered more effective in maintaining prosodic information, resulting in more expressive speech with faster inference times \cite{fastspeech1, fastspeech2}. The VITS framework is another influential E2E framework that provides waveforms as outputs \cite{vits}. Apart from eliminating the need for vocoders, this framework also removes the need for any grapheme-to-phoneme (G2P) layer and accepts raw text as input for subsequent alignment and modelling. Owing to its non-autoregressive nature, it is fast and hence suitable for low-resource contexts. Considering this, in the current work we model a baseline Mizo TTS system with Tacotron2 and then build an improved version with VITS. As Mizo is a low-resource, tonal language \cite{Lalhminghlui_Sarmah_2025}, we would like to explore how the challenges posed by data unavailability are mitigated using the two frameworks. It is worth mentioning here that while there have been some works in the acoustic-phonetics \cite{sarmah2015contextual}, tone recognition \cite{gogoi2020lexical, gogoi2021learning, gogoi25b_interspeech} and automatic speech recognition of Mizo \cite{bawitlung2025mizo}, there are no available public domain resources that demonstrate the development of a TTS system for the language. However, during 2010-11, a Mizo text reading system was developed by the Advance Signal Processing Group, Speech Processing Section, Centre for Development of Advanced Computing (C-DAC), Kolkata, India. However, no technical documentation of the system is available in the public domain \cite{cdactts}. 

Several attempts have been made to build TTS in low-resource languages in India \cite{prakash2023towards}. A TTS system was developed for Lambani, a low-resource Indo-Aryan language, with only $2,000$ sentences \cite{dasare2022text} using the Tacotron2  architecture \cite{shen2018natural}. Similarly, a low-resource TTS was developed for the Chattisgarhi language using the LRSpeech framework \cite{tts_chattisgarhi,lrspeech}. While these languages are extremely low-resource in India, even official languages such as Assamese, Bodo, Dogri, etc. suffer from insufficient speech data for TTS training \cite{tts_maha}. Several frameworks have emerged over the last 5 years for developing TTS for low-resource languages. A zero-shot multi-speaker TTS model, namely, SC-GlowTTS was proposed, which improves similarity for speakers not seen during training \cite{casanova2021sc}. Furthermore, zero-shot learning methods for TTS development have also been attempted previously \cite{fujita2024lightweight, casanova2024xtts, wang2024investigation, palatucci2009zero}. As mentioned earlier, the LRSpeech framework was specifically developed for facilitating the development of low-resource TTS systems. Considering these developments, we describe the development of a low-resource TTS for Mizo, with a focus on accurate tone synthesis.


The rest of the paper is arranged as follows. Section \ref{sec:method} describes the methodology of the current work. This section provides background information on the Mizo language and details the data collection, annotation, and post-processing. The training and testing strategies, as well as the subjective and objective evaluation of TTS outputs, are also discussed. The following Section \ref{sec:results} provides a description of the results and discussion, and finally, Section \ref{sec:conclude} concludes the paper.

\section{Methodology}
\label{sec:method}
Mizo is a South Central Tibeto-Burman language of the Kuki-Chin subfamily, and it is spoken broadly in Mizoram, a Northeastern state of India, and approximately more than half a million people in Mizoram and its neighbouring states, such as Assam, Manipur, Meghalaya, and Tripura, speak Mizo \cite{sarmah2010preliminary, sarma2018robust}. Mizo is also spoken in parts of neighbouring countries to India, including in Bangladesh and in Myanmar \cite{sarmah2010preliminary, lalhminghlui2018production}. Mizo is a low-resource language, and the Mizo orthography is based on the Roman script \cite{mizo-reading}. Like several other Kuki-Chin languages, Mizo is tonal and exhibits phonetic peculiarities, such as voiceless sonorants and a lexical tone system \cite{wendyphd}. Mizo has four lexical tones, namely High, Low, Rising and Falling \cite{fanai2015tones, gogoi2020lexical}. However, in the Mizo orthography, these tones are not consistently marked. Considering the language's phonetic peculiarities, it is interesting to see how a TTS system performs in it. 

Mizo is a low-resource language; therefore, it is not possible to find a speech database for developing TTS in the system. Considering this, the current work collected speech data from a female Mizo speaker, with the text drawn primarily from various online sources. The Mizo speaker recorded a total of $10.01$ hours of Mizo speech, of which $5.18$ hours were curated and used to model the TTS reported in the current work. 

\subsection{Recording procedure}
\label{sec:database}
The recordings for the Mizo speech database were conducted in the sound-attenuated booth of the Phonetics and Phonology Laboratory of the Indian Institute of Technology Guwahati. A female Mizo speaker recorded 10.01 hours of data over 5 days, with sessions ranging from 30 to 60 minutes. The speaker recorded the speech continuously, often in a narration or news-reading style. The speech data was recorded on a Tascam DR-100 MKII recorder using a Shure SM-10A microphone. The data was recorded with a sampling frequency of $44.1$ kHz in $16$-bit resolution. Once a session was over, the speech data was transferred to a laptop computer for annotation and further processing.

\subsection{Annotation and tokenization}
\begin{figure}[ht]
    \centering
    \includegraphics[width=\linewidth]{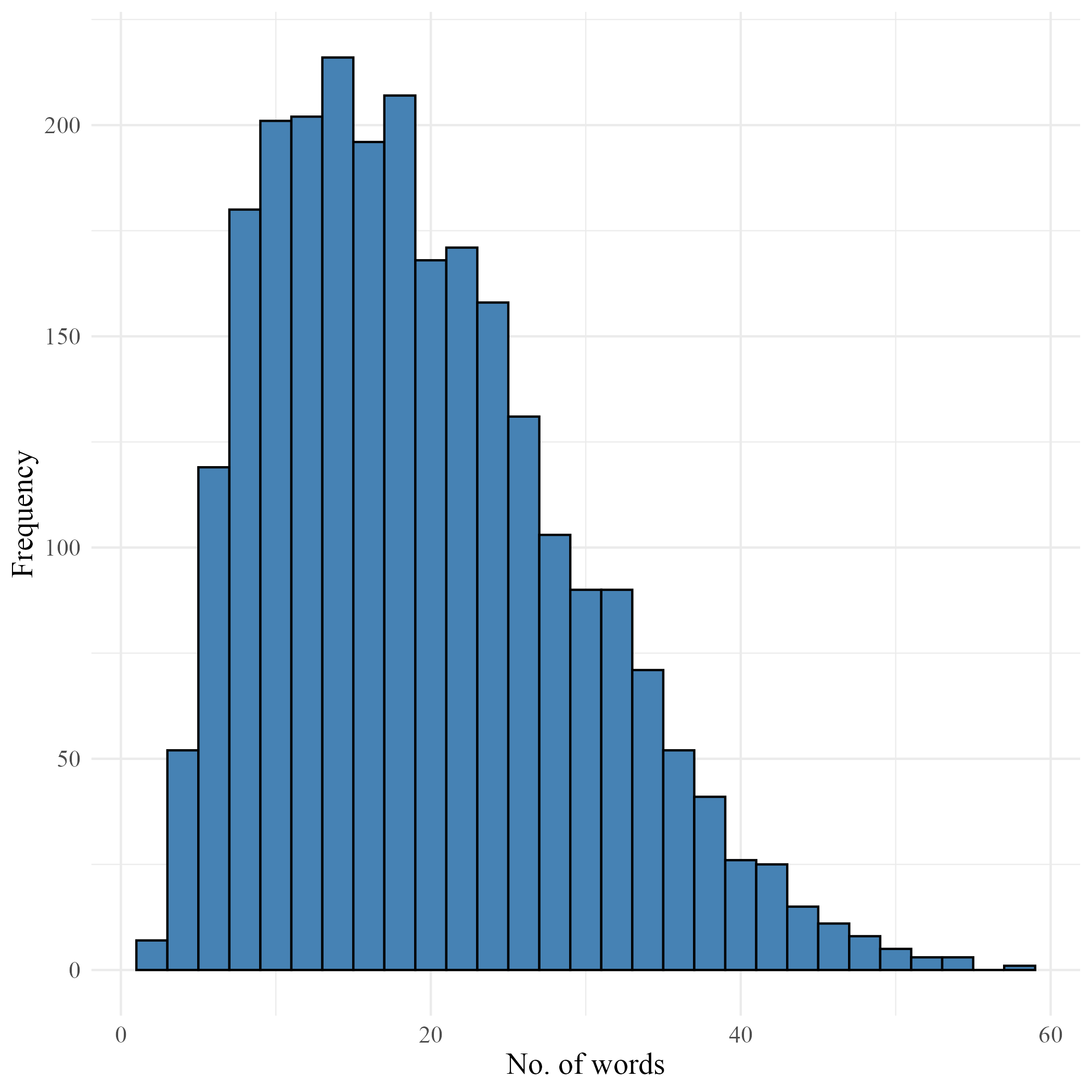}
    \caption{Histogram showing frequency of sentences with respect to the number of words contained in them.}
    \label{fig:mizo-hist}
\end{figure}

An experienced annotator segmented the speech, which was in several passages, into sentence-level by marking the boundaries using Praat 6.4.09 \cite{boersma2001praat}. The sentence-level text annotations were saved in the Praat TextGrid format. Later, a script was written to extract sentences and separate them into speech files at the sentence level, along with their annotations. In the annotations, the Mizo spelling convention was maintained. For the $5.18$ hours of data used for training and validation, the total number of sentences was $2,252$. Apart from that $25$ sentences were held out for subjective evaluation, and was never used in training. For each sentence, the file naming convention is MZ000XXX-YY, where the first five characters are the language identifier, XXX is the number associated with the paragraph from which the text was read, and YY is the exact number of the sentence in the paragraph XXX. As shown in Table \ref{tab:mizo-db}, the total number of unique words (including derived words) is $6,646$. Table \ref{tab:mizo-db} provides several other statistics of the Mizo speech database used in the current work. The number of words per sentence averaged about $22.37$, and a histogram with the distribution of the sentences by the number of words in them is provided in Figure \ref{fig:mizo-hist}.

\begin{table}[]
\setlength{\tabcolsep}{12pt} 
\caption{Statistics of the speech and text resources in training and validation.}
\begin{tabular}{lr}
\hline
\textbf{Attribute}                             & \textbf{Value} \\ \hline
Number of sentences                    & 2,252           \\
Number of words                        & 50,368          \\
Number of unique words                 & 6,646           \\
Minimum words in a sentence        & 01             \\
Maximum words in a sentence        & 59             \\
Average number of words per sentence & 22.37          \\
Total duration of the database (in hours)    & 5.18           \\
Average duration of sentence (in seconds)    & 8.28           \\ \hline
\end{tabular}

 \label{tab:mizo-db}
\end{table}

\subsection{Post-processing of audio and text data}
To develop the TTS system for the Mizo language, we utilised our own recorded database, as described in the previous section. The TTS models are designed utilising the ESPnet2 toolkit \cite{espnet}. Tacotron2 and the VITS models are natively supported in the ESPnet2 toolkit. To make the training of the TTS models effective, we had to undertake post-processing of both the speech and corresponding text data. Firstly, the speech samples were downsampled to 22.05 kHz, allowing for faster modelling without significantly compromising audio quality. Secondly, we wrote a Python script to convert numerals to words in Mizo. The code for the same is provided as supplementary material with this paper. Thirdly, the special characters with no phonetic correspondents were removed, and those with phonetic correspondents were converted to words using a custom Python script. Fourthly, abbreviations, such as Mr., Rev., Dr., etc., were expanded to ‘Mister’, ‘Reverend’, ‘Doctor’, etc. This ensured that there is one-to-one correspondence between the text and the speech files, which in turn would make the TTS system more effective. Once the post-processing was over, the sound file names were written to a comma-separated values (.csv) file with corresponding text transcriptions of the sound files. This constituted the metadata for TTS model training.

\subsection{Training strategies} \label{Training strategies}
Two neural text-to-speech architectures were investigated in this study: Tacotron2 and VITS. All models were implemented using the ESPnet2 speech processing toolkit. For Tacotron2, mel-spectrograms predicted by the acoustic model were converted into waveforms using a Parallel WaveGAN neural vocoder trained on the same Mizo speech corpus. VITS directly generates waveform outputs without the need for an external vocoder. The Tacotron2-based system follows the conventional two-stage TTS pipeline, consisting of an acoustic model and a neural vocoder. In contrast, VITS is an end-to-end architecture that jointly models acoustic representation learning and waveform generation via variational inference and adversarial training.

We trained a Tacotron2 text-to-speech model using the corpus described in Section \ref{sec:database}. The model uses 512-dimensional character/phoneme embeddings, a convolutional encoder with three layers (512 channels, kernel size 5) followed by a single 512-unit BLSTM, and a location-sensitive attention mechanism with cumulative attention. The decoder consists of two 1024-unit LSTM layers with a two-layer prenet (256 units), operating at a reduction factor of 1, and a five-layer convolutional postnet. Guided attention loss ($\sigma$ = 0.4, $\lambda$ = 1.0) and a stop-token loss were employed to stabilize alignment learning. Training used the Adam optimizer (learning rate of $1 \times 10^{-3}$), gradient clipping at 1.0, and length-based batching. 

Experiments were performed utilizing two NVIDIA V100 SXM2 GPU cards and an Intel Xeon Skylake 6248 processor. The GPU cards have 16 GB of second-generation high-bandwidth memory (HBM2). Subsequently, we trained a VITS end-to-end text-to-waveform text-to-speech model using ESPnet2 on a single-speaker speech corpus sampled at 22.05 kHz. While no modifications were made to the VITS and Tacotron2 architectures, parameters were tuned according to the requirements. The generator employs a Conformer-based text encoder with six blocks and 192 hidden channels, a posterior encoder with 16 convolutional layers, normalizing flows for latent modelling, and a HiFi-GAN–based decoder for waveform generation. A multi-scale multi-period discriminator was used for adversarial training. The training objective combines adversarial, Mel-spectrogram, feature-matching, duration, and KL-divergence losses. Both the generator and discriminator were optimised using AdamW with a learning rate of $2 \times 10^{-4}$ and exponential learning-rate decay. Training was performed for up to 500 epochs using length-based batching on two GPU nodes (each with 16 GB of memory), and the best models were selected based on training progress.

Of the $2,252$ sentences in our database, $250$ were used for validation, and the remaining $ 2,002$ sentences were used exclusively for training. The $25$ sentences used for objective and subjective evaluation, described in Section \ref{sec:eval}, are not part of these sentences.

\subsection{Objective and subjective evaluation}
\label{sec:eval}
In order to conduct objective and subjective analyses on the TTS outputs, $25$ sentences were held out. These sentences were recorded by the Mizo speaker. However, these $25$ sentences never occurred in any training or validation set. The same $25$ sentences were generated using the Tacotron2 and VITS models and included in the evaluation set for subjective and objective analyses. This resulted in an evaluation set consisting of $75$ sentences. For objective evaluation, the recent speech quality metric, known as Deep Noise Suppression Mean Opinion Score (DNSMOS) was used. The metric is modelled from the scores from ITU-T Rec. P.808 subjective evaluation \cite{itu-t}. This metric yields scores that strongly correlate with human evaluators' ratings \cite{dnsmos}. A Python script was used to extract DNSMOS scores from the 75 utterances mentioned previously. While DNSMOS scores are indicative of the perceptual quality of the synthesized speech, they are unable to specifically focus on the correctness of the phoneme realization and tonal prosody. On the other hand, considering the unique phonemes in Mizo and the presence of lexical tones in the language, we believe that fine-grained phonetic and prosodic objective evaluation is imperative for Mizo TTS-generated speech. Hence, in addition to DNSMOS, the Mel Cepstral Distortion (MCD) measure was used to quantify the spectral similarity between the TTS output and the original speech. 

MCD measures the mean difference in Mel-frequency cepstral coefficients (MFCCs) between the original speech and the synthesized speech. Hence, it is considered a good objective measure for segmental reproduction in TTS-generated speech \cite{mcd}. However, MCD is agnostic to prosodic differences, which is crucial because the Mizo TTS should reproduce the specific F0 patterns associated with the language's lexical tones. Hence, Root Mean Squared Error (RMSE\_$f0$), indicating the distance between the F0 contour of the original speech and the synthesized speech, is computed on the time axis \cite{rmse_hermes95_eurospeech, rmse_clark99_eurospeech}. Along with RMSE\_$f0$, F0 correlation (F0\_corr) coefficients are also computed, which indicate whether the direction of the F0 contours is correlated or not \cite{rmse_clark99_eurospeech}. For DNSMOS, MCD, RMSE\_$f0$, and F0\_corr, the scores were also subjected to statistical analysis to determine significance. Considering the observations are limited to $25$ in each group, paired \textit{t-tests} \cite{ttest} were conducted to see the significance of the difference between the Tacotron2 and VITS outputs. 

In the subjective evaluation, $35$ Mizo native speakers participated ($27$ female and $8$ male). The evaluators had to respond to whether the sound samples were synthesised (`Artificial') or natural human speech (`Real'). Subsequently, they were asked to rate the speech samples on a 5-point Likert scale, where 5 was `Excellent', 4 was `Good', 3 was `Fair or OK', 2 was `Poor', and 1 was `Bad'. The evaluation was designed as a web application that the evaluators could access online. After the evaluators provided their ratings and assessments, the data were collated into a comma-separated values (.csv) file. However, only $2601$ observations could be recorded, as $24$ responses were not recorded by the web interface. The .csv file was read into R-core \cite{R-core} and subject normalization was conducted to remove idiosyncratic biases. The \textit{z-score} normalization was conducted \cite{andrade2021z}, and later the values were rescaled to MOS-like values by multiplying the average \textit{z-score} values by the global standard deviation of the data and adding the global average to them. 

Following this, the data were analysed using R \cite{R-core}, and descriptive and inferential statistics were conducted. The objective results were compared for statistical significance with \textit{t-test} \cite{t-test} using the \textit{doBy} package on R \cite{doBy}. The subjective MOS results were modeled using the Linear Mixed Effects (LME) model using the \textit{lme4} package on R \cite{lme4}. The model was further subjected to a post hoc Bonferroni test for pairwise comparisons of MOS scores across natural, Tacotron2, and VITS outputs using the \textit{emmeans} package \cite{emmeans}. Finally, the \textit{ggplot2} package in R was used for data visualization \cite{ggplot2}.

\section{Results and discussion}
\label{sec:results}
\subsection{Objective evaluation}
\begin{table}[]
    \centering
    \setlength{\tabcolsep}{3pt} 
    \caption{Results of DNSMOS, MCD, RMSE\_$f0$, F0\_corr, and MOS scores evaluating original and synthesized speech from Tacotron2 and VITS TTS models.}
    \begin{tabular}{lccc}
    \hline
       Metric  & Natural Speech & Tacotron2 & VITS \\
       \hline
        DNSMOS↑&3.84&3.85&3.88\\
        MCD (dB)↓&NA&2.53&2.32\\
        RMSE\_$f0$ (Hz)↓&NA&41.34&40.60\\	
        F0\_corr ↑&NA&0.24&0.25\\
        MOS (raw)↑&4.13&2.68&3.21\\
        MOS (scaled)↑&4.18&2.63&3.46\\
    \hline
    \end{tabular}
    
    \label{tab:evalresults}
\end{table}

The results of the objective evaluation on $25$ sentences are provided in Table \ref{tab:evalresults}. The VITS system achieved numerically better results than Tacotron2 across all four objective metrics. However, the difference was statistically significant only for MCD, indicating lower spectral distortion for VITS. The differences in DNSMOS, RMSE\_$f0$ and F0 correlation were not statistically significant. The DNSMOS scores for VITS and natural speech are comparable, $3.88$ and $3.84$, respectively. VITS achieved a slightly higher DNSMOS score than Tacotron2, although the difference was not statistically significant. These scores provide an indication of overall speech quality but do not establish equivalence to natural speech. At the same time, the MCD values are lower for the VITS outputs, indicating there is better spectral similarity between the ground truth (natural human speech) and the synthesized outputs. This suggests that, at the segmental level, the VITS outputs maintain near-real phonetic similarity in terms of vowels and consonants. The results of the RMSE\_$f0$ and F0\_corr showed that the outputs of both Tacotron2 and VITS models were not highly correlated with the ground truth. However, the VITS system still outperformed Tacotron2 in terms of numerical results. This indicates that the outputs still require improvement in terms of prosody.

To determine if the outputs of the Tacotron2 and VITS models are significantly different, paired t-tests were conducted on the DNSMOS, MCD, RMSE\_$f0$, and F0\_corr scores of the $25$ outputs from each model. The results of the t-test are summarised in Table \ref{tab:ttest}. The difference between the VITS and the Tacotron2 model is statistically significant (\textit{p}$<$0.0001), in terms of MCD, with the VITS model outperforming the other. However, in terms of the DNSMOS results, RMSE\_$f0$ and F0\_corr, the results are not significantly different (\textit{p}$>$0.05), even though the VITS model marginally outperforms the Tacotron2 model.

\begin{table}[]
    \centering
    \setlength{\tabcolsep}{4.5pt} 
    \caption{Results of the t-tests conducted for DNSMOS, MCD, RMSE\_$f0$, F0\_corr values to see the difference (VITS - Tacotron2) of the outputs from the Tacotron2 and VITS models.}
    \begin{tabular}{lcccc}
    \hline
       Metric  & Mean difference & \textit{t-value}&df&\textit{p-value} \\
       \hline
        DNSMOS&0.03&0.36&24&0.72\\
        MCD (dB)&-0.21&-8.35&24&$<$0.0001\\
        RMSE\_$f0$ (Hz)&-0.74&-0.52&24&0.61\\	
        F0\_corr &0.01&0.43&24&0.67\\
    \hline
    \end{tabular}
    \label{tab:ttest}
\end{table}

\subsection{Subjective evaluation}
\label{sec:subj_eval}
Subjective evaluation was conducted using the MOS score paradigm and based on native Mizo speakers' judgments of the naturalness of the TTS outputs. As shown in Table \ref{tab:evalresults}, the raw MOS scores and the normalized and scaled MOS scores show that the outputs from the VITS models are considered better in terms of speech quality by the native speakers of Mizo.

\begin{figure}[]
    \centering
    \includegraphics[width=\linewidth]{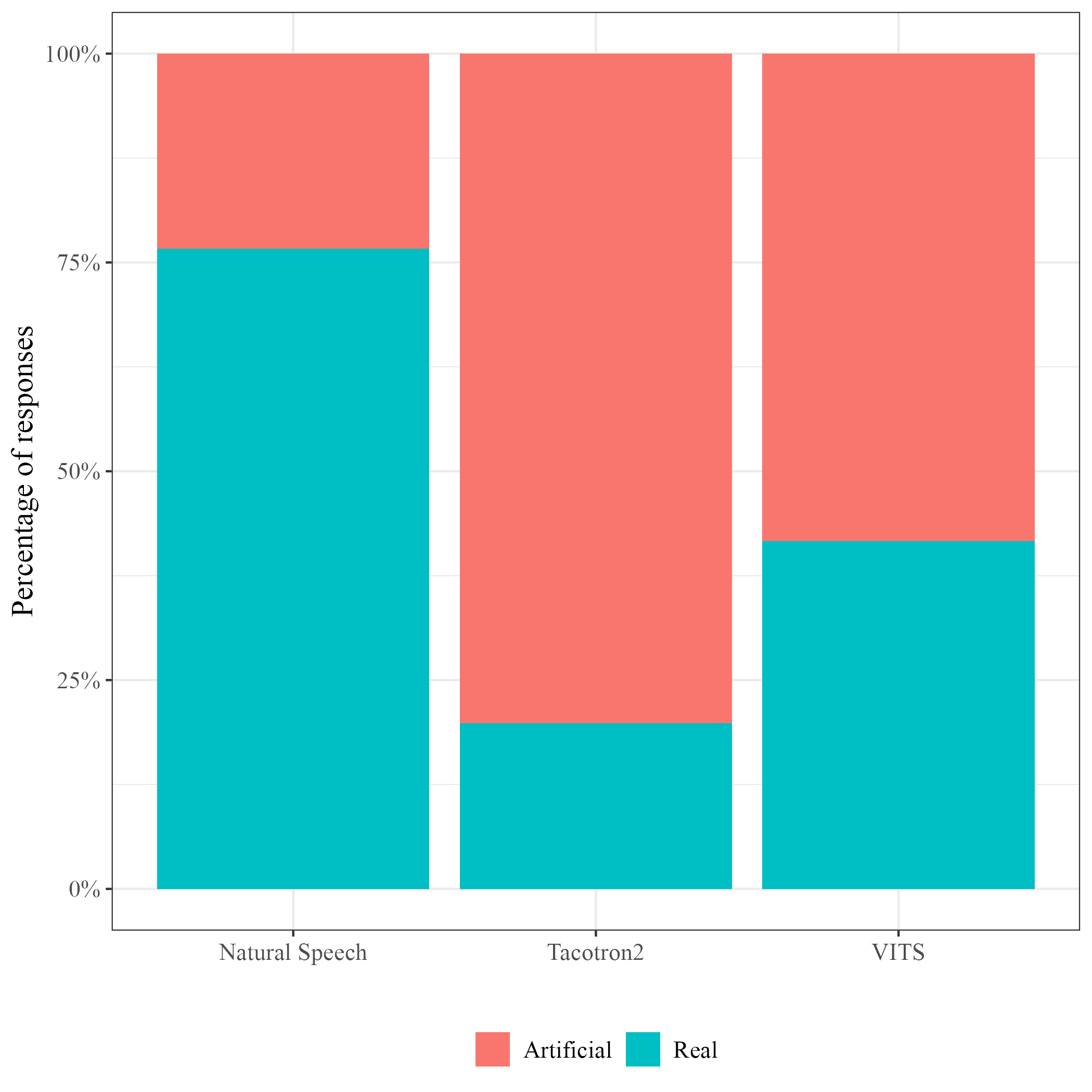}
    \caption{Human evaluators' judgment of original and synthesized speech samples in terms of their naturalness.}
    \label{fig:realarti}
\end{figure}

We also examined the naturalness ratings of the original speech sounds and the TTS-generated outputs provided by the native speakers. The overall trend is shown in Figure \ref{fig:realarti}, where $76\%$ of the sentences in the ground truth, i.e. natural speech, are considered natural by the native speakers. For Tacotron2 and VITS-generated sentences, the percentage of sentences that were considered natural speech is $20\%$ and $41\%$, respectively. This suggests that, despite low RMSE\_$f0$ and F0\_corr scores, native speakers of Mizo consider the TTS outputs, specifically those from VITS, to be of good perceptual quality. 

\begin{figure}[ht]
    \centering
    \includegraphics[width=\linewidth]{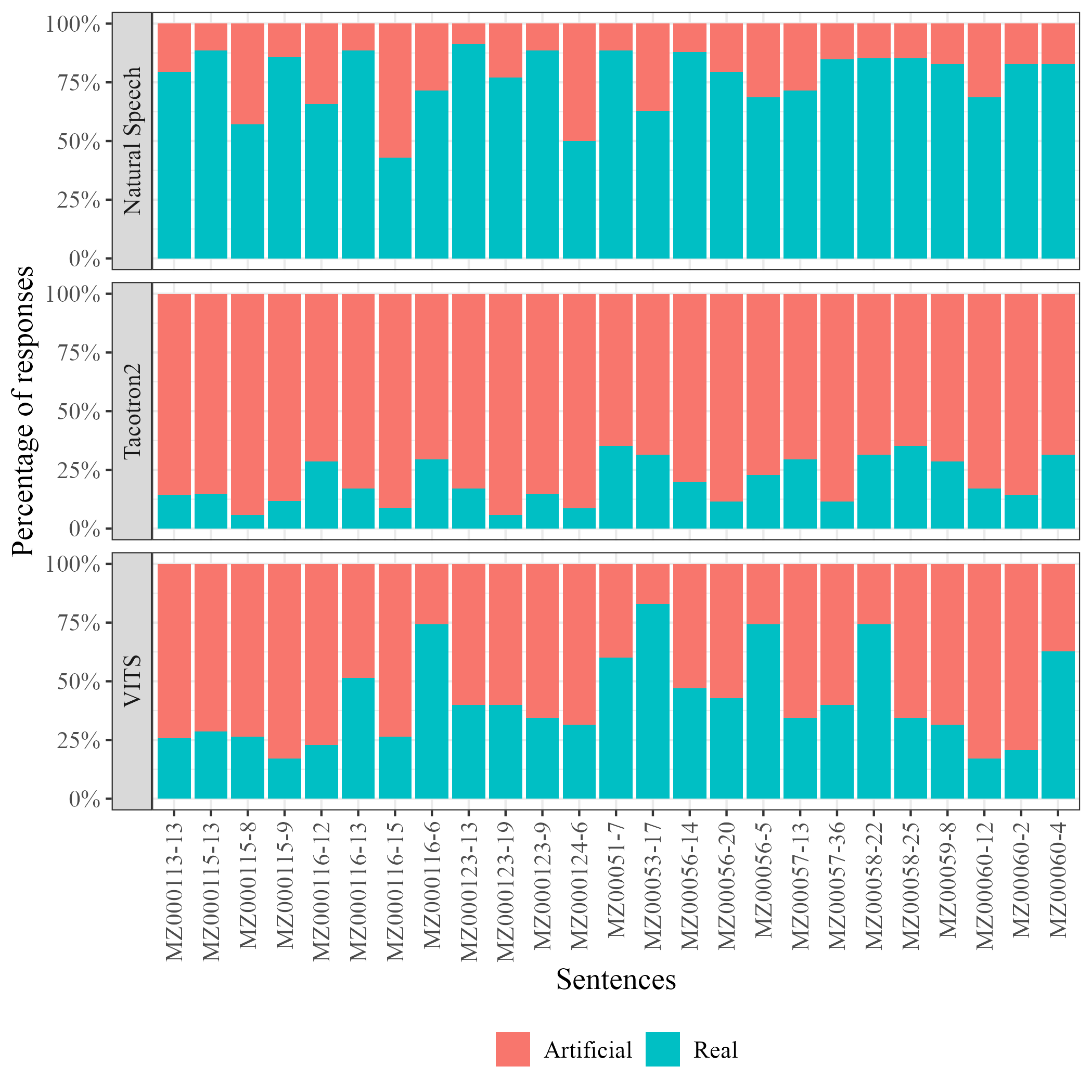}
    \caption{Sentence-wise representation of human evaluators' judgment of original and synthesized speech samples in terms of their naturalness.}
    \label{fig:realarti-sent}
\end{figure}

Sentence-wise naturalness ratings provided by Mizo native speakers were also examined. The sentence-wise percentage of ratings given by human evaluators to natural and synthesized speech is shown in Figure \ref{fig:realarti-sent}. The figure shows that, broadly speaking, the trends for the Tacotron2 a- and VITS-generated outputs are similar; however, in absolute terms, the VITS outputs are higher than those of Tacotron2. In three sentences, the Tacotron2-generated outputs are marginally considered more natural than the VITS-generated ones. These are sentences MZ00058-25, MZ00060-12 and MZ000116-25, as seen in Figure \ref{fig:realarti-sent}.

For inferential statistics, because the dataset of MOS scores is larger (i.e., containing $2601$ observations in total), we conducted an LME test \cite{lme4} to assess whether the MOS scores differ significantly. The model used was as in Equation \ref{eq:lme}, where $Type$ consists of ratings for natural speech, Tacotron2, and VITS. $Sentence$ is a unique sentence in the text, and the Subjects are the individual raters. In the model, $Type$, $Sentences$ and their interaction are considered fixed effects, and $Subject$ is considered the random effect. 

\begin{equation}
\label{eq:lme}
    MOS \sim Type + Sentence + Type\times Sentence + (1|Subject)
\end{equation}

\begin{table}[ht]
    \centering
    \setlength{\tabcolsep}{12pt}
    \caption{Results of Type II Wald $\chi^2$ test on the LME model.}
    \begin{tabular}{lccc}
    
    \hline
       &$\chi^2$&\textit{df}&\textit{p}-value\\ 
    \hline
       Type&1233.68& 2 &$<$0.001\\
       Sentence&241.72&24&$<$0.001\\
       Type$\times$ Sentence&223.04&48&$<$0.001\\
    \hline
    \end{tabular}
    
    \label{tab:lmeresults}
\end{table}

\begin{table}[]
    \centering
    
    \setlength{\tabcolsep}{3pt} 
     \caption{Pairwise comparison of MOS scores using Bonferroni post-hoc.}
    \begin{tabular}{lccccc}
    \hline
Pair&Estimate&SE&df&\textit{t}-ratio&\textit{p}-value\\
\hline
 Natural - Tacotron2&1.45&0.04&2642&34.19&$<$0.0001\\
 Natural - VITS&0.93&0.04&2642&21.84&$<$0.0001\\
 Tacotron2 - VITS&-0.53&0.04&2642&-12.38&$<$0.0001\\
 \hline
    \end{tabular}
   
    \label{tab:pairwise}
\end{table}

\begin{table}[h]
    \centering
    \setlength{\tabcolsep}{9pt} 
     \caption{Tone error rates (TER) in \% for Tacotron2 and VITS models, the final row shows mean sentence-level TERs.}
    \begin{tabular}{lccc}
    
\hline
Sentence ID&No. of TBU&Tacotron2&VITS\\
\hline
MZ000113-13	&	25	&	28.00	&	16.00	\\
MZ000115-13	&	22	&	13.64	&	9.09	\\
MZ000115-8	&	30	&	10.00	&	3.33	\\
MZ000115-9	&	22	&	18.18	&	13.64	\\
MZ000116-12	&	25	&	4.00	&	12.00	\\
MZ000116-13	&	16	&	6.25	&	0	\\
MZ000116-15	&	18	&	11.11	&	5.56	\\
MZ000116-6	&	23	&	21.74	&	0	\\
MZ000123-13	&	13	&	15.38	&	0	\\
MZ000123-19	&	18	&	11.11	&	0	\\
MZ000123-9	&	18	&	33.33	&	0	\\
MZ000124-6	&	20	&	15.00	&	15.00	\\
MZ00051-7	&	20	&	5.00	&	5.00	\\
MZ00053-17	&	15	&	13.33	&	0	\\
MZ00056-14	&	29	&	10.34	&	10.34	\\
MZ00056-20	&	23	&	13.04	&	0	\\
MZ00056-5	&	22	&	13.64	&	4.55	\\
MZ00057-13	&	16	&	6.25	&	12.50	\\
MZ00057-36	&	15	&	20.00	&	6.67	\\
MZ00058-22	&	15	&	13.33	&	0	\\
MZ00058-25	&	18	&	5.56	&	11.11	\\
MZ00059-8	&	24	&	12.50	&	0	\\
MZ00060-12	&	18	&	11.11	&	5.56	\\
MZ00060-2	&	18	&	5.56	&	5.56	\\
MZ00060-4	&	17	&	5.88	&	5.88	\\
\hline
\textbf{Mean TER:}&&\textbf{12.93}&\textbf{5.67}\\

 \hline
    \end{tabular}
   
    \label{tab:ter}
\end{table}

    

\begin{table}[]
    \centering
    \setlength{\tabcolsep}{12pt} 
    
 \caption{Distribution of errors in percentage of total errors by Mizo tone categories.}
    \begin{tabular}{lcc}
    \hline
Tones&Tacotron2 (n = 65)&VITS (n = 30)\\
\hline
High	&	33.9\%	    &	33.3\%\\
Low	    &	40.0\%	    &	40.0\% \\
Rising	&	4.6\%	    &	13.3\% \\
Falling	&	21.5\%	    &	13.3\%\\
\hline
\end{tabular}
    \label{tab:errors}
\end{table}

After fitting the linear mixed-effects model (LME), Type II Wald chi-square tests were performed using the Anova() function from the R package \textit{car} to assess the statistical significance of the fixed effects. The results are presented in Table \ref{tab:lmeresults}. The results indicate that the three sets of audio are judged significantly differently by the Mizo speakers, as evidenced by their MOS scores. Apart from that, the significant effect of sentences also indicates that the MOS ratings vary across sentences. Finally, the significant interaction between the audio type and sentences indicates that the effect of audio type on MOS ratings varies across sentences.

To conduct a pairwise comparison of the MOS scores provided for the three audio types, namely, natural human speech, Tacotron2-generated and VITS-generated, we subjected the LME model to a pairwise comparison using the Bonferroni post-hoc method with the \textit{emmeans} package in R. The results of the pairwise comparison are shown in Table \ref{tab:pairwise}. The table shows that the MOS values differed significantly, with the VITS system outperforming Tacotron2 by $0.53$. At the same time, the sentences produced by the speaker (Natural category) were rated higher than both Tacotron2 and VITS outputs. All the pairwise comparisons were statistically significant ($p~< 0.001$).

\subsection{Tone Error Rates (TER) in synthesized outputs}
The Tacotron2 and VITS outputs were evaluated by an experienced phonetician who is also a native speaker of the Mizo language. The evaluator was asked to indicate the syllables on which the models generated the tone incorrectly. The detailed feedback from the evaluator was quantified and presented in Table \ref{tab:ter}. Overall, the VITS system performs better at synthesizing the Mizo tones. In Table \ref{tab:ter}, the tone error rates (TER) are calculated by taking the percentage of wrongly produced tones in the total number of potential tone-bearing units (TBU) in a Mizo sentence \cite{wendyphd}. As shown in the table, the mean sentence-level TER for the Tacotron2-derived synthesis is approximately $12.93\%$, which decreases to $5.67\%$ in the VITS-derived system. As each TBU has a tone by default in the TTS outputs, the errors are substitution errors. As shown in Table \ref{tab:errors}, the error trend is similar for Tacotron2 and VITS-based systems. The highest number of errors is attributed to the incorrect synthesis of the `High' and `Low' tones in Mizo. Nevertheless, we do not think the system finds synthesizing `High' and `Low' tones challenging. When looking into the distribution of the tones of Mizo, naturally occurring in the language, we notice that the `High' and the `Low' tones are more common than the `Rising' and the `Falling' tones \cite{wendyphd}. Hence, there is more probability of the occurrence of these tones and consequently more errors may be attributed to these two categories of tones. However, it is also to be noted that the `High' and the `Low' tones are level tones (even though `'Low' tones do exibit a slight fall) and hence, there are more chances that they be misrecognized and confused with each other. However, that needs further analysis of the error produced in the Mizo speech synthesis systems. Nevertheless, the reduced tone errors in the VITS system indicate thatVITS-generated Mizo speech is more robust than that from Tacotron2. This may be the reason the subjective evaluators of Mizo synthesized speech, as shown in Section \ref{sec:subj_eval}, find the VITS-derived outputs more natural.

\section{Conclusion}
\label{sec:conclude}
This paper reported an attempt to develop a TTS model for the Mizo, a low-resource tonal language, using 5.18 hours of curated speech from a single female speaker. We compared Tacotron2, coupled with a neural vocoder, and the end-to-end VITS framework using objective metrics, subjective ratings from native speakers, and expert assessment of tone realization. There were substantial challenges, given that the language is low-resource and has a complex phoneme inventory with four distinct lexical tones. Hence, Mizo TTS development had to address for the language's phonetic nuances. Given this, two approaches were taken: the Tacotron2 and VITS frameworks. While Tacotron2 is widely used in TTS development, the VITS framework is known to enhance naturalness by incorporating variations in prosody and rhythm, owing to its use of variational autoencoders and adversarial training \cite{xie2025controllablespeechsynthesisera}. Our primary objective was to investigate how effectively the two systems are able to synthesize Mizo speech, specifically in the absence of tone-specific orthographic cues.   

The objective evaluation of the outputs showed that VITS achieved significantly lower Mel Cepstral Distortion (MCD). At the same time, it also yielded significantly higher subjective ratings than Tacotron2. However, the differences in DNSMOS, F0 RMSE, and F0 correlation were not statistically significant. While subjective evaluation showed significantly higher scores for the VITS system, not all objective measures corroborated its advantages. Hence, it shows that complementary metrics in TTS evaluation are necessary, as spectral similarity, predicted perceptual quality, and human ratings capture different aspects of synthesized speech.

One of the major focuses of this work was the evaluation of the lexical tones in the synthesized Mizo speech. Expert evaluation of 25 held-out sentences also showed a lower mean sentence-level tone error rate for VITS (5.67\%) than for Tacotron2 (12.93\%). While both systems were trained without explicit tone markings, the VITS system performed better at reproducing Mizo tones compared to Tacotron2. This also may suggest that language-specific tone distribution patterns and contextual information may have been exploited by the model to predict and synthesize the lexical tones, even when explicit tone markings are absent. While we are sceptical of making such a claim at this point in time, carefully curated diverse speech data accompanied by text may be sufficient for modeling tones in low-resource tonal languages without the need for explicit tone marking, reducing the need for tedious tone annotation in the databases.


As the rest of the Mizo data is already curated, a natural next step for this work will be to assess the effect of training data quantity in improving synthesis quality. Evaluating the models against the systems reported here will help determine whether increased training data improves spectral similarity, perceived naturalness, and tonal accuracy, and whether these improvements are consistent across these dimensions. In future, we also intend to include additional speakers and examine how speaker diversity affects synthesis quality and tonal realization, specifically in low-resource languages. Additionally, exploration of different speaking styles, including teaching, news reading, and storytelling, will support the development of more flexible TTS systems for practical applications while also contributing to a better understanding of data-efficient prosodic modelling.

\section{Data and samples}
The training data is not available publicly. To listen to the samples, please visit {\href{https://www.iitg.ac.in/priyankoo/mizo/mizo-tts/play.php}{https://www.iitg.ac.in/priyankoo/mizo/mizo-tts/play.php}}. For a live demonstration of the Mizo TTS, please visit {\href{https://clst.iitg.ac.in/tts/mizo-tts/}{https://clst.iitg.ac.in/tts/mizo-tts/}}.

\section{Acknowledgment}
The initial preparations for training the Mizo TTS system would not have been possible without the suggestions from Dr Antti Suni of the University of Helsinki and Dr Sishir Kalita and Dr Anirban Dutta of Armsoftech.AIR. This work is supported by the grant ``Speech Technologies for North Eastern Languages”,
awarded to the third and fourth authors by the Ministry of Electronics and Information Technology (MeitY), Government of India, under the NLTM Bhashini project.

\bibliographystyle{IEEEtran}
\bibliography{mybib}

\end{document}